# Methodology proposal for proactive detection of network anomalies in e-learning system during the COVID-19 scenario

Ivan Cvitić[0000-0003-3728-6711], Dragan Peraković[1][0000-0002-0476-9373], Marko Periša[1][0000-0002-1775-0735] and Anca D. Jurcut[2][0000-0002-2705-1823]

[1] University of Zagreb, Faculty of Transport and Traffic Sciences, Vukelićeva 4, 10000 Zagreb, Croatia
`ivan.cvitic@fpz.unizg.hr, dragan.perakovic@fpz.unizg.hr,`
`marko.perisa@fpz.unizg.hr`
[2] University College Dublin, School of Computer Science and Informatics Belfield, Dublin 4, Ireland
`anca.jurcut@ucd.ie`

**Abstract.** In specific conditions and crisis situations such as the pandemic of coronavirus (SARS-CoV-2), or the COVID-19 disease, e-learning systems became crucial for the smooth performing of teaching and other educational processes. In such scenarios, the availability of e-learning ecosystem elements is further highlighted. An indicator of the importance for securing the availability of such an ecosystem is evident from the DDoS (Distributed Denial of Service) attack on AAI@EduHr as a key authentication service for number of e-learning users in Republic of Croatia. In doing so, numerous users (teachers/students/administrators) were prevented from implementing and participating in the planned teaching process. Given that DDoS as an anomaly of network traffic has been identified as one of the key threats to the e-learning ecosystem in crisis scenarios, this research will focus on overview of methodology for developing a model for proactive detection of DDoS traffic. The challenge in detection is to effectively differentiate the increased traffic intensity and service requests caused by legitimate user activity (flash crowd) from the illegitimate traffic caused by a DDoS attack. The DDoS traffic detection model developed by following analyzed methodology would serve as a basis for providing further guidelines and recommendations in the form of response to events that may negatively affect the availability of e-learning ecosystem elements such as DDoS attack.

## 1   Introduction

Anomaly represents samples in the data that deviate from the previously defined normal behavior of the observed phenomenon. Observed from the aspect of information and communication system, anomalies in communication, i.e. network traffic are generally generated by one or several network devices. This is often a result from illegitimate network activities in the system, with the anomalies of network traffic having the potential of negatively affecting the operation of the information and communication system or services [1]. One of the frequent causes of anomalies in the network traffic is DDoS (Distributed Denial of Service) attack. Over the last two decades numerous studies have been directed to the development of methods, models and systems that can detect DDoS traffic in real time. Nevertheless, the number of DDoS attacks and the amount of DDoS traffic is constantly increasing, which is the reason for further research in the area of the detection of security threats of this kind [2]. Despite continuous research into network traffic anomalies, cyber-attacks such as DDoS attacks are still frequent and can have numerous negative effects on the predicted performance of IC (Information and Communication) systems and the availability of IC services. The pandemic of coronavirus (SARS-CoV-2) highlighted the importance of the availability of e-learning systems and services. Goal of this research is to propose research methodology for development of DDoS traffic detection models at the attack target in scenario where flash events generated traffic represents legitimate traffic, such as e-learning services during COVID-19 pandemic.

## 2    Previous research

Previous researches define several approaches to DDoS traffic detection. Generally, they can be divided into two basic categories, based on the samples and based on the anomalies [3]. Research [4], apart from the previous ones, identifies also the approach based on entropy, and research [5] mentions the possibilities of applying the hybrid approach to DDoS traffic detection. The methods based on the sample apply the comparison of the incoming traffic with the pre-defined profiles and samples of the known network anomalies [6]. The detection of DDoS attack based on the sample can be carried out in three ways; based on the signature of the known attacks, based on the rule (if-then) and based on the condition and transition [7]. The advantage of this method of detection is a high rate of detection of the already known DDoS attacks with a low number of false positive and false negative results. The drawback is the impossibility to detect new and unknown attacks, i.e. those attacks that are not included in the database which is used for the comparison with the samples of the incoming traffic. Because of the dynamics of the problem area, it is of great importance that the detection methods be able to detect unknown samples of DDoS traffic [3].
On the contrary, the approach based on the detection of the network traffic anomalies uses the pre-defined models of normal traffic which are then used to compare the incoming traffic [5]. This approach to detection has been developed in order to overcome the drawbacks of the detection approach based on the samples [4]. If the incoming traffic significantly differs from the defined model of normal traffic, then the incoming traffic is identified as anomaly, i.e. as DDoS traffic [8]. The advantage of detection of network traffic anomalies related to the detection based on samples is the possibility of discovering unknown attacks. The main drawback of detection based on anomalies is the problem of determining the threshold values between normal traffic and anomaly [5], [9]. The anomalies of network traffic are detected when the values of the current traffic flow or other selected parameters exceed the pre-defined thresh-old of the normal traffic model. A low defined threshold results in a large number of false positive results, and high defined threshold leads to a large number of false negative results [10]. Numerous scientific methods have been used for the detection of DDoS traffic [11]. The current academic literature most frequently applies the statistical methods, machine learning methods and softcomputing methods [12], [13], [14]. In today's big data concept environments, high performances in cyber-attack detection and response are demonstrated by machine learning and artificial intelligence methods.

## 3 Methodology for DDoS traffic detection model development

Despite continuous research into network traffic anomalies, cyber-attacks such as DDoS attacks are still frequent and can have numerous negative effects on the predicted performance of IC systems and the availability of IC services. The pandemic of coronavirus (SARS-CoV-2) highlighted the importance of the availability of e-learning systems and services. Crisis situations such as the mentioned pandemic result in the need for isolation of users (students/teachers/administrators) whereby education and supporting processes rely on the reliable work of the e-learning system and all his elements. On March 16th, 2020 there was a DDoS attack on AAI@EduHr system responsible for authenticating users to access various e-learning services (Merlin, webinar, e-learning center, *Dabar*, *Hrčak*, filesender and others) in Republic of Croatia [15]. The conducted attack indicated the need to research and find solutions for cyber threats (primarily DDoS) that are applicable in specific scenarios.

According to data from the Croatian University Computing Center (*Srce*) during March 2020, in the month in which remote education began due to the COVID-19 pandemic, the AAI@EduHr system recorded 11,214,236 successful authentications for 517,453 unique users (Fig. 1.). In comparison, in March 2019, the AAI@EduHr system had a total of 3,220,212 successful authentications of 252,974 unique users, which can be seen in Fig. 2. Authentication through AAI@EduHr was most commonly used to access MS Office 365 systems for schools, Loomen, and *Srce* systems: Merlin - a remote learning system for students and teachers, and the ISVU Information System for Higher Education (primary the *Studomat* module) [16].

The presented data indicate the occurrence of the flash crowd phenomenon. This occurs when legitimate requests to access a web-based service exceed the statistically normal number of legitimate requests [17]. Accordingly, flash crowds may adversely affect the performance of DDoS attack detection models based on machine learning methods that use data sets created during the period when the number of service requests is common. Such models are often detecting flash crowd traffic as a DDoS attack traffic, even though it represents a legitimate service request. Some authors are researching solutions that can differentiate flash crowd and DDoS traffic by using human behavior and interaction, but such solutions are not acceptable to users [18].

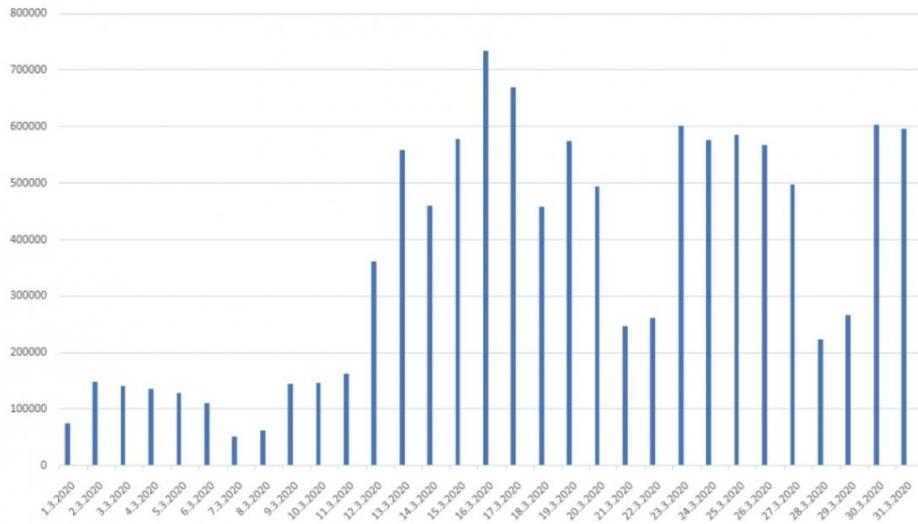

**Fig. 1.** Number of successful authentications in March 2020 by day [16]

In order to achieve effective detection of DDoS and thus for reaction to such attacks in specific cases of flash crowd phenomena under crisis situations such as a coronavirus pandemic, the detection model should be based on the ability to distinguish between DDoS traffic and traffic generated under flash crowd conditions. One of research direction followed by this project proposal is to identify the unique characteristic of flash crowd traffic vis-à-vis DDoS traffic as a basis for development of DDoS traffic detection model [19].

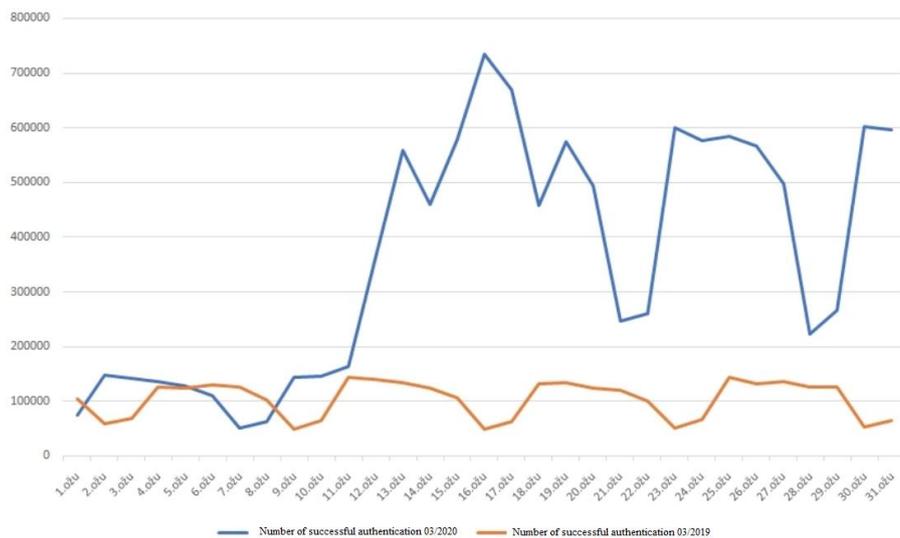

**Fig. 2.** Number of successful authentications March 2019 and 2020 comparation [16]

Implementation of efficient and effective research for dealing with described problem requires well-structured research methodology.

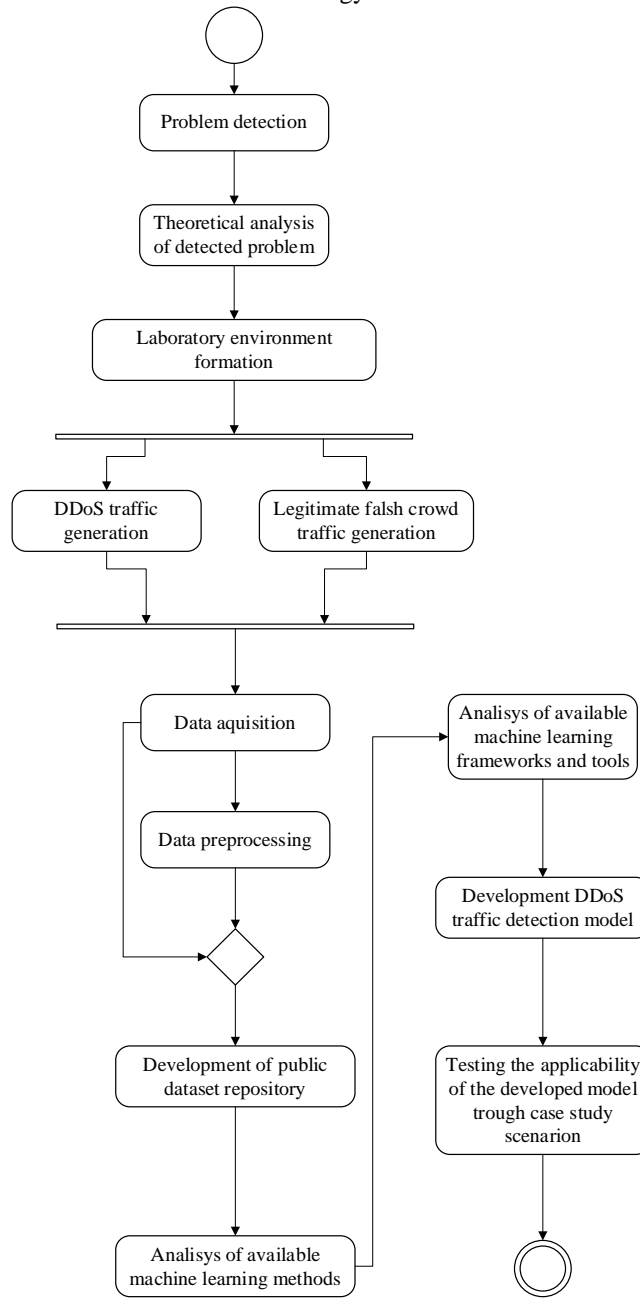

**Fig. 3.** Activities through research phases

We propose research implementation through four phase methodology, and activities shown in Fig. 3. as follows: (1) analysis of existing elements of the e-learning ecosystem and establishing the theoretical basis and differences between DDoS and flash crowd traffic, (2) forming of a laboratory environment, the collection of network traffic, processing and preparation of collected data for further analysis, (3) development of DDoS traffic detection model, validation and performance evaluation of the developed DDoS traffic detection model, (4) analysis of the applicability of the developed anomaly detection model and the reaction capabilities based on the operation of the developed model.

### 3.1  Establishing the theoretical basis and differences between DDoS and flash crowd traffic

In the first phase of the research the current scientific literature will analyze for the purpose of identifying the elements of the e-learning ecosystem, architecture, communication topologies and technologies, and other relevant characteristics of such environments. The purpose of the aforementioned research activities is to provide adequate recommendations and guidelines in the last phase of the research with the purpose of ensuring the availability of e-learning services and minimizing the negative impacts of the emergence and realization of cyber threats such as DDoS attacks. The analysis of the current scientific literature will also determine the current research findings related to the characteristics of network traffic generated as a result of flash crowd activity and current research achievements regarding the development of models and systems that can distinguish DDoS and flash crowd traffic.

### 3.2  Formation of laboratory environment and data collection

Second phase should include establishing a laboratory environment for generating and collecting legitimate (flash crowd) and illegitimate (DDoS) traffic. The laboratory environment is planned to be established at the Department of Information and Communication Traffic at the FPZ within the Laboratory for Security and Forensic Analysis of the Information and Communication System (LSF). Flash crowd and DDoS traffic generation is planned with the implementation of the dedicated hardware platform. Such a platform is able to simulate realistic flash crowd and DDoS traffic and also provides advanced management and defining features of the generated traffic at different layers of the OSI (The Open Systems Interconnection) model. The next planned activity is to collect and store the generated traffic on the dedicated computers in .pcap format files, which is suitable for further manipulation in the form of analysis and extraction of the traffic features values. Further research activities in this phase are relate to the pre-processing and preparation of data for further analysis. This implies the filtering of the collected data and extraction of the network traffic features that can be use in network anomaly detection model and adequate labeling of feature vectors. Considered will be exclusively the packet header values, i.e. statistical features of traffic while the packet content will not be considered because of the application of cryptographic methods in the communication processes. In this phase, the independent traffic features will be selected based on which it is possible to differentiate the observed traffic on legitimate and DDoS. Also, the level of connection with the dependent feature will be determined and those independent features with the highest level of connection with the dependent feature will be selected. The selected features will be used for the purpose of defining the model in the next research phase. Final output of second phase will be datasets containing legitimate traffic generated as a result of flash crowd activities on e-learning ecosystem elements (simulated trough dedicated network generator hardware) and DDoS traffic as an network traffic anomaly generated as a result of DDoS attack (simulated trough dedicated network traffic generator hardware). By collecting generated legitimate and illegitimate traffic, a unique dataset will be formed. Such dataset will be published on a publicly available server (on FPZ's servers) in original and in the processed form for academic community with purpose of further research of detection and reaction on DDoS attacks in specific scenarios.

### 3.3 Development of DDoS traffic detection model

Previously carried out activities laid the foundation for the third phase of planed research. This phase encompasses the development of the network traffic anomalies detection model. The first step in this phase is to analyze and select adequate supervised machine learning method from the set of ensemble methods that will suit the solving of binary classification problem. In solving such a problem, the objective is to check the congruence of the generated new traffic sample with the sample of the legitimate traffic. The incongruity of the traffic features values with the values of the features of legitimate traffic above the defined threshold will mean that the device generates DDoS traffic within the observed time. Based on previous mentioned, following activity is development of network anomaly detection model using supervised machine learning methods. For implementation of chosen method available machine learning platform will be used such as TensorFlow, Weka, KNIME, Orange, Hadoop, Apache Spark, Neo4ji, R or similar. The last step of this phase is to validate the model performances through standard validation measures for classification models (accuracy, precision, specificity, kappa coefficient, rate of false and true positive results, confusion matrix, etc.).

### 3.4 Applicability of the developed anomaly detection model

In the final, fourth research phase the applicability of the developed model in real scenario through case study will be analyzed. Also, the guidelines and recommendations for response to the anomalies detected with the developed model will be defined. Purpose of this activity is minimizing the negative effects of cyberthreats such as DDoS on availability of e-learning services particularly in crisis situations such as pandemic of SARS-CoV-2 virus when such services are critical for the undisturbedly running of educational and supporting processes.

## 4     Discussion and conclusion

Using proposed methodology for developing model of network traffic anomaly detection as well as guidelines and recommendations for reaction to detected anomalies provide the potential for further practical application, as well as strong socio-economic benefits from several aspects. The research implementation within the framework of the proposed project is significant for the development of the research area since it considers the challenges in a specific scenario of using e-learning services resulted from COVID-19 pandemic. In proposed research it is planned to form open and public data set repositories containing network traffic (in raw and preprocessed format) generated simulating flash crowd scenarios and DDoS attacks. Such repository will benefit other researchers for the purpose of further research of behaving flash crowd traffic, how to distinguish it from other types of traffic such as DDoS or some other type of traffic anomalies. Planed extensive use of machine learning, especially ensemble type of machine learning methods will potentially result with developed anomaly detection model that can proactively detect cyber-threat and adequately react on such a threat. In that way it will give possibility to secure availability of e-learning ecosystem and its services when it is needed the most, and that is in the crisis scenarios such as pandemic of coronavirus and similar. Securing availability of e-learning services that are becoming critical will increase user satisfaction on every level (students/teachers/administrators) and allow undisturbedly running of educational and supporting processes. Furthermore, this kind of research would open variety of related research problems and it would enable knowledge transfer between divers' research teams and e-learning system operators as a response to high-risk situations and crisis scenarios that may occur in the future.

There are several results that are expected trough implementation of proposed methodology. First expected result is determined theoretical basis of characteristics and main differences between traffic generated during flash crowd events and DDoS attacks discovered by other researchers. Second expected result is formed laboratory environment at LSF for the purpose of generating legitimate and illegitimate network traffic. Third expected result is collected dataset of legitimate and illegitimate traffic adequately labelled and prepared for further analysis and research. Fourth expected result is repository of datasets of collected legitimate and illegitimate traffic available for further research to academic community. Fifth expected result is developed network traffic anomaly detection model that is based on ensemble methods of supervised machine learning. Sixth expected result is defined guidelines and recommendation for reaction on detected cyberthreats for minimizing negative impact on availability of e-learning services in crisis scenarios (such as COVID-19 pandemic) when such services are crucial in effective implementation of the educational and supporting processes.

# References

1. Aggarwal CC. Outlier Analysis. Artificial Intelligence Review. 2017;24(2):379–84. Available from http://link.springer.com/10.1007/978-3-319-18123-3%0A
2. Deka RK, Bhattacharyya DK. Self-similarity based DDoS attack detection using Hurst parameter. Security and Communication Networks. 2016;9(17):4468–81.
3. Tan Z, Jamdagni A, He X, Member S, Nanda P, Member S, et al. Detection of Denial-of-Service Attacks Based on Computer Vision Techniques. IEEE Transactions on Computers. 2015;64(9):1–14.
4. David J, Thomas C. DDoS Attack Detection Using Fast Entropy Approach on Flow-Based Network Traffic. Procedia Computer Science. 2015;50:30–6. Available from http://linkinghub.elsevier.com/retrieve/pii/S1877050915005086
5. Mirkovic J, Reiher P. A taxonomy of DDoS attack and DDoS defense mechanisms. ACM SIGCOMM Computer Communication Review. 2004;34(2):39. Available from http://portal.acm.org/citation.cfm?doid=997150.997156
6. Bhuyan MH, Bhattacharyya DK, Kalita JK. Network Anomaly Detection: Methods, Systems and Tools. IEEE Communications Surveys & Tutorials. 2014;16(1):303–36.
7. Bhattacharyya DK, Kalita JK. DDoS Attacks: Evolution, Detection, Prevention, Reaction and Tolerance. Boca Raton, USA: CRC Press; 2016.
8. Zeb K, AsSadhan B, Al-Muhtadi J, Alshebeili S. Anomaly detection using Wavelet-based estimation of LRD in packet and byte count of control traffic. In: Proceedings of 7th International Conference on Information and Communication Systems (ICICS). IEEE; 2016. p. 316–21.
9. Xiang Y, Li K, Zhou W. Low-rate DDoS attacks detection and traceback by using new information metrics. IEEE Transactions on Information Forensics and Security. 2011;6(2):426–37.
10. Zargar ST, Joshi J, Tipper D. A survey of defense mechanisms against distributed denial of service (DDOS) flooding attacks. IEEE Communications Surveys and Tutorials. 2013;15(4):2046–69.
11. Cvitić I, Peraković D, Periša M, Husnjak S. An Overview of Distributed Denial of Service Traffic Detection Approaches. PROMET - Traffic&Transportation. 2019;31(4):453–64. Available from https://traffic.fpz.hr/index.php/PROMTT/article/view/3082
12. Bhuyan MH, Kashyap HJ, Bhattacharyya DK, Kalita JK. Detecting Distributed Denial of Service Attacks: Methods, Tools and Future Directions. The Computer Journal. 2014;57(4):537–56.
13. Cvitić I, Peraković D, Periša M, Botica M. Novel approach for detection of IoT generated DDoS traffic. Wireless Networks. 2019;25:1–4.
14. Peraković D, Periša M, Cvitić I, Husnjak S. Model for detection and classification of DDoS traffic based on artificial neural network. Telfor Journal. 2017;9(1):26-31.
15. SRCE. Povećano opterećenje sustava AAI@EduHr | Srce [Internet]. Available from https://www.srce.unizg.hr/vijesti/povecano-opterecenje-sustava-aaieduhr/objav2020-03-16 [accessed 2020 Apr 4]
16. SRCE. Rekordan broj autentikacija putem SSO servisa sustava AAI@EduHr | Srce [Internet]. Available from https://www.srce.unizg.hr/vijesti/rekordan-broj-autentikacija-putem-sso-servisa-sustava-aaieduhr/objav2020-04-02 [accessed 2020 Apr 4]
17. Behal S, Kumar K, Sachdeva M. Discriminating flash events from DDoS attacks: A comprehensive review. International Journal of Network Security. 2017;19(5):734–41.

18. Yu S, Zhou W, Jia W, Guo S, Xiang Y, Tang F. Discriminating DDoS attacks from flash crowds using flow correlation coefficient. IEEE Transactions on Parallel and Distributed Systems. 2012;23(6):1073–80.
19. Bhandari A, Sangal AL, Kumar K. Characterizing flash events and distributed denial-of-service attacks: an empirical investigation. Security and Communication Networks. 2016;9:2222–39. Available from http://doi.wiley.com/10.1002/sec.1472